\documentclass[aps,
superscriptaddress,
twocolumn,
twoside,
floatfix,
pra,
nofootinbib,
10pt,
a4paper]{revtex4-2}
\usepackage{graphicx} 

\usepackage{dcolumn}
\usepackage{bm}
\usepackage[dvipsnames]{xcolor}
\begin{document}
\preprint{APS/123-QED}

\title{Memory resistor based in GaAs 2D-bilayers: In and out of equilibrium}

\author{C. Marty}
\affiliation{Laboratory for Solid State Physics, ETH Zürich, CH-8093 Zürich, Switzerland}
\affiliation{Quantum Center, ETH Zürich, CH-8093 Zürich, Switzerland}

\author{Z. Lei}
\affiliation{Laboratory for Solid State Physics, ETH Zürich, CH-8093 Zürich, Switzerland}
\affiliation{Quantum Center, ETH Zürich, CH-8093 Zürich, Switzerland}

\author{S. Silletta}
\affiliation{Laboratory for Solid State Physics, ETH Zürich, CH-8093 Zürich, Switzerland}
\affiliation{Quantum Center, ETH Zürich, CH-8093 Zürich, Switzerland}

\author{C. Reichl}
\affiliation{Laboratory for Solid State Physics, ETH Zürich, CH-8093 Zürich, Switzerland}
\affiliation{Quantum Center, ETH Zürich, CH-8093 Zürich, Switzerland}

\author{W. Dietsche}
\affiliation{Laboratory for Solid State Physics, ETH Zürich, CH-8093 Zürich, Switzerland}
\affiliation{Quantum Center, ETH Zürich, CH-8093 Zürich, Switzerland}
\affiliation{Max-Plank-Institut für Festkörperforschung, D-70569 Stuttgart, Germany}

\author{W. Wegscheider}
\affiliation{Laboratory for Solid State Physics, ETH Zürich, CH-8093 Zürich, Switzerland}
\affiliation{Quantum Center, ETH Zürich, CH-8093 Zürich, Switzerland}


\begin{abstract}
    Resonant tunneling between closely spaced two dimensional electron gases is a single particle phenomenon that has sparked interest for decades. High tunneling conductances at equal electron densities are observed whenever the Fermi levels of the two quantum wells align. Detuning the Fermi levels out of the resonant 2D-2D tunneling regime causes a negative differential resistance. The negative differential resistance leads to a hysteresis when operating the device in a current driven mode, allowing a bilayer system to function as a volatile memory resistor. 
\end{abstract}
\maketitle
Equilibrium resonant electron tunneling is a well-established single-particle phenomenon in density-balanced bilayer two-dimensional electron gases (2DEGs). This tunneling process has been successfully modelled by considering the conservation of energy and in-plane electron momentum~\cite{Murphy1995,Brown1995,Simmons1993,Zheng1993}. The tunneling current can be derived using Fermi's golden rule for electron-electron interactions in bilayer tunneling systems~\cite{Zheng1993}. Recent studies on bilayer systems such as bilayered Graphene-h-BN, Graphene-WSe$_2$, WSe$_2$ and GaAs bilayers have revived interest in density-balanced tunneling~\cite{Britnell2013,Burg2017,Burg2018, Shi2022, Marty2023}, demonstrating significantly enhanced interlayer tunneling.

Here, we investigate the transport characteristics of electron bilayers, consisting of two GaAs quantum wells (QWs). These are separated by Al(Ga)As barriers of exceptionally narrow widths that have not been systematically investigated so far. A method of epitaxially integrating patterned back gates developed in our group, allow to establish ohmic contacts to each QW independently. The tunneling conductivities in our devices agree remarkably well with the single particle resonant tunneling (RT) model. As expected, current-voltage dependent (I-V) characteristics show an augmented interlayer conductivity at balanced densities and a negative differential resistance (NDR) for interlayer voltages going beyond balanced densities.

However, under forced tunneling currents, the voltage-current dependent (V-I) curve features distinct discontinuities. The 2D-2D tunneling conductivity slowly decreases from its maximal value for increasing tunneling currents, until a critical tunneling current is reached. At this point, a pronounced interlayer voltage build-up immediately misaligns the Fermi levels of the 2DEGs, abruptly reducing the absolute tunneling conductivity. Furthermore, the V-I trace is hysteretic due to the NDR behavior of the device. This hysteretic dependence on the tunneling current enables the device to function as a hysteretic threshold current switch, also known as volatile memory resistors, which have been investigated in a variety of material systems~\cite{Slesazeck2019,Wang2020,Kim2022,Zuo2023}.

Additionally, we show that the 2D-2D RT is highly susceptible to a perpendicular magnetic field and temperature as previously reported for RT~\cite{Simmons1993,Aleiner1995,Turner1996}. Specifically, we show that the critical current for the current switch is reduced by the magnetic field and the step-like increase in conductivity becomes progressively smoother. 

\begin{figure}
    \centering
    \includegraphics[width=86mm]{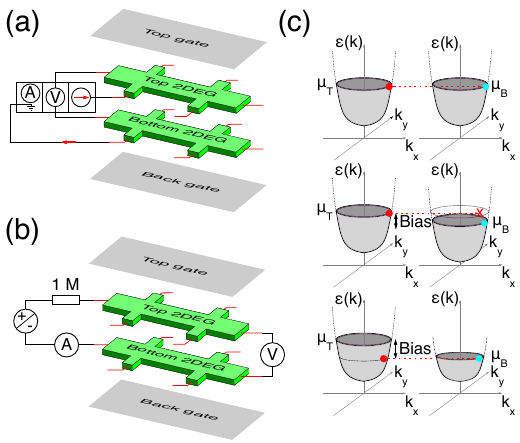}
    \caption{\textbf{(a)} Schematic of the bilayered Hall bars. The active measurement scheme is depicted, in which the voltage is controlled and the current is measured. \textbf{(b)} The passive measurement forces a tunneling current and the corresponding interlayer voltage is measured. \textbf{(c)} Dispersion relations for both electron layers. In the top panel the densities are balanced and there is no voltage bias applied between the layers. In the center panel a voltage bias misaligns the two Fermi levels. In the bottom panel the densities are imbalanced and there is a voltage bias between the two layers.}
    \label{fig: resonnant_tunneling}
\end{figure}
Our devices are molecular beam epitaxially grown on GaAs wafers. Device \textbf{A} has two 18.7 nm wide GaAs QWs and an Al$_{0.8}$Ga$_{0.2}$As barrier of 6 nm, similar to the device investigated in~\cite{Marty2023}. Device \textbf{B} has two 25 nm wide GaAs QWs and an Al$_{0.8}$Ga$_{0.2}$As barrier of 8 nm.
The heterostructure was designed to achieve balanced densities in the QWs and high charge carrier mobilities, utilizing an optimized design of large setback distances and asymmetric $\delta$-layers of Si-dopants above and below the QW region. Device A exhibits a mobility of  \mbox{$\approx1.8\cdot10^6\textit{ }$ cm$^2$V$^{-1}$s$^{-1}$} at a density of $0.92\cdot10^{11}\textit{ }$ cm$^{-2}$ in each QW. Device B has a mobility of  $\approx2\cdot10^6\textit{ }$ cm$^2$V$^{-1}$s$^{-1}$ at a density of $1.3\cdot10^{11}\textit{ }$ cm$^{-2}$ in each QW.

A Hall bar, 200 $\mu$m wide and 1250 $\mu$m long, was fabricated by photolithography and wet-chemical etching. A patterned, conducting, silicon doped GaAs epilayer with local oxygen ion implantation forms pinch-off back gates, allowing to contact the two layers independently~\cite{Berl2016a, Scharnetzky2020}. Similarly, Hall bar spanning global top and back gates allow to tune the electron densities of each 2DEG independently. 
\begin{figure}[t!]
    \centering
    \includegraphics[width=86mm,keepaspectratio]{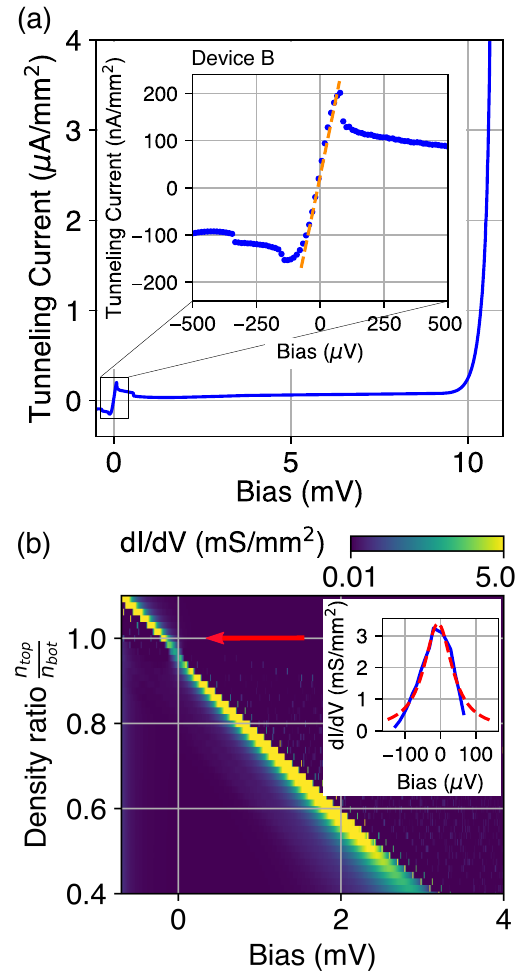}
    \caption{\textbf{(a)} Tunneling current as a function of interlayer bias for device B acquired by the active measurement. RT occurs near zero bias. Above 10 mV the tunneling current increases strongly. Inset: Resonance at balanced densities near zero interlayer bias (dots). Using the model from eq. \ref{eg: tunneling_current} the calculated tunneling current is displayed for \mbox{$\Delta_{SAS} = 140\textit{ }\mu$eV} and $\Gamma = 48.7\textit{ }\mu$eV (dashed). \textbf{(b)} dI/dV tunneling conductance as a function of the density ratio n$_{top}$/n$_{bot}$ and interlayer voltage bias of device B. The tunnel conductance is increased whenever the Fermi levels of the two QWs align and momenta matched states are available. Inset: dI/dV of the resonant I-V curve depicted in (a) near zero bias (solid). A Lorentzian is fitted to the dI/dV. The fitting parameters are \mbox{$\omega_0 = -7.03$ $\mu$eV} and the full-width at half-max is \mbox{$2\Gamma=97.4$ $\mu$eV} (dashed).}
    \label{fig: voltage_bias}
\end{figure}

We record the interlayer tunneling conductivity at cryogenic temperatures in the mK range in a dilution refrigerator, using two different methods: An active and a passive measurement, depicted in Fig. \ref{fig: resonnant_tunneling}(a) and (b). For the active method, a voltage regulator controls the interlayer voltage via a 4-point measurement. The regulator pushes and pulls current to and from each of the 2DEGs independently until the desired interlayer bias is reached. The tunneling conductivity is calculated from the interlayer bias and the tunneling current provided by the regulator. The second method is passive, whereby a current is forced to tunnel between the two 2DEGs and the resulting interlayer voltage drop between the layers is recorded. The tunneling current is generated by a source measure unit with a 1 M$\Omega$ preresistor. The tunneling conductance is calculated from the measured interlayer voltage and the current drawn from the source unit. 

In Fig. \ref{fig: voltage_bias} we display I-V and dI/dV data from device B using the voltage regulator. In (a) the gate voltages are set such that the densities in the two 2DEGs are balanced, n$_{top} =$ n$_{bot} = 1.3\cdot10^{11}\textit{ }$ cm$^{-2}$. In the inset we sweep the interlayer voltage from -0.5 mV to \mbox{0.5 mV} using the regulator and record the tunneling current. Whenever the Fermi levels of the two QWs align, the rate of change for the tunneling current becomes maximal, resulting in a Lorentzian peak for the conductivity and a peak tunneling conductivity of up to 3.25 mS/mm$^2$, as seen in the inset of (b).

In the off-resonant tunneling regime the conductance remains in the low microsiemens range. Notable is the interlayer bias region around the steep slope in Fig. \ref{fig: voltage_bias}(a). Here the tunneling current drops and exhibits a negative differential behavior. The sudden reduction in tunneling current and conductivity stems from the misalignment of the Fermi levels between the two 2DEGs provoked by the interlayer bias. In (b) we show the dI/dV for different ratios of top to bottom layer densities. The conductivity peak shifts linearly with the change in the density ratio.
\begin{figure*}[t]
    \centering
    \includegraphics[]{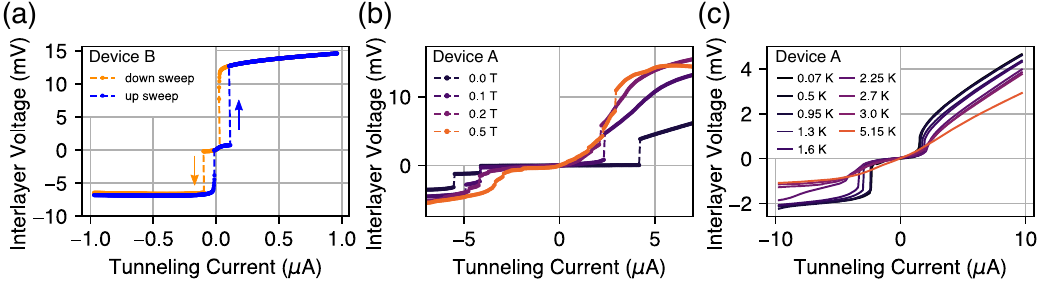}
    \caption{\textbf{(a)} Forced current measurements for two sweep directions on device B. 2D-2D RT for small currents around 0 nA. At a critical current the interlayer voltage jumps by several millivolts. There is a hysteresis when changing the sweep direction. \textbf{(b)} The interlayer voltages for forced tunneling currents for different perpendicular magnetic fields for device A are depicted. \textbf{(c)} Increasing the temperature softens the voltage discontinuity until it vanishes at high temperatures.}
    \label{fig: forced_tunneling}
\end{figure*}

The calculation of the RT current is based on Fermi's golden rule. An electron can tunnel between the layers only when the state it occupies in one QW is unoccupied in the other, i.e. the energy and the in-plane momentum of the electron must be conserved. The conservation of both properties strongly limits the available states. 
2D-2D tunneling occurs in the narrow region for balanced densities and zero interlayer bias where the Fermi levels of the two 2DEGs align and whenever the finite interlayer bias voltage compensates for the imbalanced densities.
This is schematically illustrated in Fig. \ref{fig: resonnant_tunneling}(c). Zheng and MacDonald~\cite{Zheng1993} derived the tunneling current to be 
\begin{equation}
    \label{eg: tunneling_current}
    I = -e\int_{-\infty}^{+\infty} T(E)[f(E-\mu_T)-f(E-\mu_B)]dE.
\end{equation}
The occupational statistics are given by the Fermi-Dirac functions of each layer. The values of the Fermi functions differ by the chemical potentials of each QW $\mu_T\textit{, }\mu_B$ (top and bottom). The chemical potential in the QW depends on the electron sheet density in the QW and the applied interlayer bias.

The tunneling probability $T(E)$ is computed by the barrier dependent energy splitting $\Delta_{SAS}$ of the symmetric and antisymmetric wave functions and a Lorentzian spectral function governing the dispersion of a quasi particle. The Lorentzian has a broadening $\Gamma$, which is equivalent to the quantum lifetime of the electrons.
\begin{equation}
    T(E) = \frac{2\pi}{\hbar}\sum_{-k_f,k_f} \Delta_{SAS}^2 \left(\frac{1}{\pi}\frac{\Gamma}{(E-\epsilon(k))^2+\Gamma^2}\right)^2
\end{equation}
Here, we use the same spectral function for both QWs. This entails the same quasi particle broadening $\Gamma$ and uses the same dispersion relation $\epsilon(k) = \hbar^2k^2/(2m^*)$. 

The calculated tunneling current at resonance agrees with the experiment, as shown in Fig. \ref{fig: voltage_bias}(a). The full-width at half-max in the inset in Fig. \ref{fig: voltage_bias}(b) determines the quasi particle broadening $2\Gamma =$ 97.4 $\mu$eV for our calculation. This value is obtained by fitting a Lorentzian to the dI/dV curve. Our $\Gamma =48.7$ $\mu$eV is small compared to reported values by Murphy et al.~\cite{Murphy1995}. The short quantum lifetimes may be a result of the high RT conductivity in our devices with thin barriers.

The remaining parameter for the calculation of the tunneling current is $\Delta_{SAS}$. A tunnel splitting of $\Delta_{SAS} = 140 \textit{ }\mu$eV fits the measured tunneling current best. This value agrees with the expected value for a bilayer system with an 8 nm barrier, where the energy level splitting between the symmetric and antisymmetric wave functions is calculated quantum mechanically~\cite{Boebinger1990}.

For the passive measurement, a constant current is forced to tunnel between the two density balanced layers. For small currents the interlayer voltage remains in the RT regime with a high conductivity and hence a small interlayer voltage, as seen in Fig. \ref{fig: forced_tunneling}(a). At a critical current of $\approx 200$ nA the interlayer voltage jumps discontinuously to several millivolts. This behavior originates from the finite tunneling conductance which, in combination with the tunneling current, yields an interlayer voltage between the layers, misaligning the Fermi levels. At an interlayer voltage of $\approx 150\textit{ }\mu$V the Fermi levels are sufficiently misaligned such that the conductivity decreases, leading to the discontinuity in the interlayer voltage. This point defines the critical current where the RT breaks down.
On the other hand, starting from a high tunneling current the Fermi levels are initially misaligned and an interlayer voltage in the millivolt regime is measured. This high interlayer voltage persists for tunneling currents below the critical current due to the NDR, which causes a direction dependent hysteresis. The hysteresis is symmetric around 0 tunneling current.

\begin{figure}[t!]
    \centering
    \includegraphics[width=86mm]{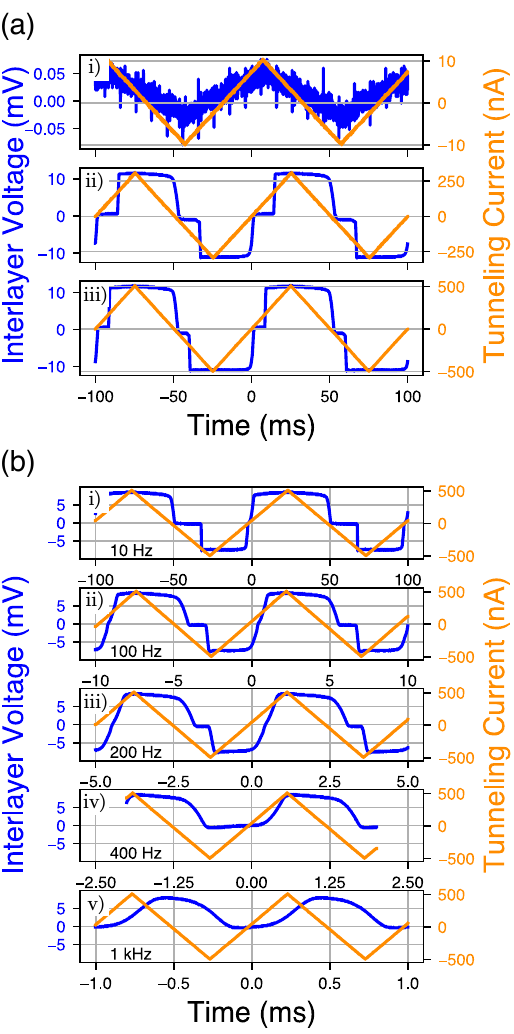}
    \caption{Interlayer voltages for passive measurements on device B. \textbf{(a)} i) For small currents the tunneling conductance is linear. The tunneling conductance originates solely from the 2D-2D RT. ii)-iii) For currents larger than the threshold current the interlayer voltage jumps. Due to the hysteresis the voltage remains high until the current returns back to zero. \textbf{(b)} Frequency dependence of the device. iv)-v) The discontinuity vanishes at high frequencies.}
    \label{fig: hysteresis}
\end{figure}

The symmetric behavior of the device is also seen in Fig. \ref{fig: hysteresis}(a) i)-iii) for time dependence. Panel i) shows the linear current voltage behavior below threshold. In this regime the interlayer conductivity is high, corresponding to RT. The hysteresis is observed once the tunneling current surpasses the threshold current, as seen in panel ii)-iii). The interlayer voltage abruptly changes and remains in the high millivolt range until the tunneling current reaches zero. The difference between panel ii) and iii) is that the critical current is reached after a shorter time.

In Fig. \ref{fig: forced_tunneling}(b) we investigate the interlayer voltage for forced tunneling currents in device A, subjected to perpendicular magnetic fields. We observe a strong magnetic field dependence. The critical tunneling current of $\approx \pm4 \textit{ }\mu A$ at zero field is first reduced and vanishes for stronger magnetic fields. For increasing magnetic fields the curve becomes continuous. The suppression of RT for increasing magnetic fields appears to be gradual, with no critical magnetic field $B_c$. This suppression of RT is due to the formation of a gap in the tunneling density of states, until inter Landau-Level RT occurs~\cite{Eisenstein1992c,Simmons1993,Brown1995, Turner1996, Reker2002}.

In Fig. \ref{fig: forced_tunneling}(c) the temperature dependence of the tunneling conductance at balanced densities for device A is shown for forced tunneling currents. Increasing the temperature from 70 mK to 0.95 K has little effect on the interlayer voltage. For temperatures above 1.3 K the critical current increases while the step height decreases, due to broadening of the Fermi distributions. This allows for a higher interlayer voltage before the Fermi levels are misaligned. Given that conductivity in the RT regime is unchanged for higher temperatures, the increased interlayer voltage range results in an increased critical current. Furthermore, this broadening excites electrons to higher momenta and energy states. This allows them to overcome the potential barrier posed by the pinch-off gates reducing the step height~\cite{Simmons1998}. For temperatures as high as 3 K and 5 K, the discontinuity eventually vanishes. As seen, the thermal broadening smears the curve completely and the slope around zero current increases, reducing the tunneling conductance.

The threshold switching behavior and the hysteresis observed in the passive measurements is an implementation of a volatile memory resistor. Here, the absolute conductivity is dependent on the history of the tunneling current. For density balanced gate configurations the two 2DEGs function as a symmetric two terminal current-switch with the critical current being the threshold value. The RT conductivity depends on the barrier thickness. At zero bias, device A has a tunneling conductance of 400 mS/mm$^2$ and device B has a conductance of 3.25 mS/mm$^2$. These values are comparable to those reported by Blount et al.~\cite{Blount1998}. The RT conductivity also determines the critical current since the interlayer voltage equates to the tunneling current divided by the conductivity. A higher conductivity leads to a higher maximal RT current. This is clearly seen by the comparison of the device A and B with $\approx$ 4 $\mu$A and $\approx$ 200 nA, respectively. The tunneling conductivity is proportional to the active tunneling area. For our devices we use the same device geometries.

From Fig. \ref{fig: hysteresis}(a) ii) an on-off ratio of 23:1 is established. This ratio depends on the potential build-up in the layers becoming sufficiently large such that electrons can hop the potential barrier imposed by the pinch off gates ~\cite{Blount1998, Simmons1998a}. A more negative pinch-off gate voltage can influence this ratio. In Fig. \ref{fig: forced_tunneling}(a) the pinch-off gate barrier potentials have been set unevenly. Hence, the ratio for negative currents is lower than for positive. In Fig. \ref{fig: hysteresis}(a) the potential barriers imposed by the pinch-off gates have been tuned symmetrically and are roughly 10 meV.

In Fig. \ref{fig: hysteresis}(b) we show the time dependence of the device B in the passive measurement setup. In panel i)-iii) the interlayer voltage jumps as the tunneling current passes the threshold current. At frequencies of 400 Hz and 1 kHz (panel iv-v), the current threshold for negative currents is not met, resulting in a low negative interlayer voltage. We attribute this frequency dependence to be intrinsic to our device. The two layers of the device form a plate capacitor. The asymptotic behavior of the reactance of the capacitor leads to an increase in the tunneling conductivity. This increase in conductivity increases the critical current and hence the device remains in the RT regime. The measurement lines and low-pass filters built-in to the experimental setup can be excluded as cause for the observed frequency dependence, as their cutoff frequency are well above 5 KHz.

To operate the device at a higher frequency one could force larger tunneling currents or reduce the device capacitance by optimizing the device geometry. Finally, we want to remark that there is a capacitive lag of the dilution fridge and the device causing a constant delay in the interlayer voltage response relative to the input signal. Independent of the input frequency we get a phase delay of 1 ms, which can be seen in panels ii)-iv).

In conclusion, we measure RT behavior for bilayer devices with 6 and 8 nm barriers. The dI/dV tunneling conductivity matches the numerical calculations backed by Fermi's golden rule. Breaking the RT condition we observe a NDR behavior for bias voltages that exceed the range of aligned Fermi levels. For current forced measurements the NDR results in a step-like increase for the interlayer voltage and for the absolute conductivity going beyond the critical current. There is a hysteresis due to the NDR in the I-V curve, hence the bilayer system functions as a threshold current switch. For balanced 2DEG densities this volatile memory resistor operates symmetrically with respect to the sign of the tunneling current. The thickness of the barrier governs the conductivity in the linear RT regime as well as the value of the critical current. Furthermore, we show that the critical tunneling current is reduced in magnetic fields. As the tunneling conductivity decreases with magnetic field the critical current and the interlayer voltage step diminish. Similarly, the voltage step in the V-I curve vanishes for increased temperature due to the thermal broadening of the Fermi distribution that obstructs the NDR.

We acknowledge financial support from the Swiss National Science Foundation (SNSF) and the NCCR QSIT (National Center of Competence in Research -
Quantum Science and Technology).
\bibliography{PhD-bilayerTunneling}
\end{document}